# OK Computer Analysis: An Audio Corpus Study of Radiohead
Nick Collins

**Abstract**

The application of music information retrieval techniques in popular music studies has great promise. In the present work, a corpus of Radiohead songs across their career from 1992-2017 are subjected to automated audio analysis. We examine findings from a number of granularities and perspectives, including within song and between song examination of both timbral-rhythmic and harmonic features. Chronological changes include possible career spanning effects for a band's releases such as slowing tempi and reduced brightness, and the timbral markers of Radiohead's expanding approach to instrumental resources most identified with the *Kid A/Amnesiac* era. We conclude with a discussion highlighting some challenges for this approach, and the potential for a field of audio file based career analysis.

**1 Introduction**

Radiohead's music has gained scholarly attention commensurate with its wider popularity and cultural reach, from the publication of Brad Osborn's monograph on analysis[1] through earlier volumes by Marianne Letts[2] and Joseph Tate.[3] The diversity of Radiohead's musical ideas is well highlighted by Osborn,[4] and their musical interest well represented in Hesselink's discussion of *Pyramid Song*, perhaps the most rhythmically ambiguous of their oeuvre.[5] Ignoring Thom Yorke's sideline as techno DJ at Exeter University, Radiohead began as a guitar band, but introduced a wider palette of timbres in their career, with electronic sound a particularly potent side avenue.[6]

We approach Radiohead's oeuvre from the perspective of computational analysis, exploring audio from throughout their career to gain insight into their musical development. The big data era has seen an associated rise in computationally driven corpus studies, allowing for the treatment of a larger database of MIDI or audio files of much greater size than heroic historic manual annotation projects.[7] The computer is a fast and undistracted analysis tool; a domain in computer music, Music Information Retrieval (MIR),[8] has especially centred on the potential to extract musical features from corpora of sound and associated meta-data. Western popular music, with its obvious commercial implications, has often been the target music for MIR researchers.[9] MIR is objective in application, but still constrained by the initial subjective decisions of programmers setting up such systems, with their associated choices over data sets, representations, and algorithms.

---

[1] Osborn, Brad. *Everything in its right place: Analyzing Radiohead*. New York: Oxford University Press, 2017
[2] Letts, Marianne T. *Radiohead and the resistant concept album: How to disappear completely*. Indiana University Press, 2010.
[3] Tate, Joseph, ed. *The Music and Art of Radiohead*. Aldershot, Hants: Ashgate Publishing Ltd, 2005.
[4] Osborn 2017
[5] Hesselink, Nathan D. "Radiohead's "pyramid song": Ambiguity, rhythm, and participation." *Music Theory Online* 19, no. 1 (2013).
[6] Letts 2010; Collins, Nick. "Kid A, and: Amnesiac, and: Hail to the Thief." *Computer Music Journal* 28, no. 1 (2004): 73-77.
[7] Clarke, Eric and Nicholas Cook, eds. *Empirical Musicology: Aims, Methods, Prospects*. Oxford: Oxford University Press, 2004; Temperley, David, and Leigh VanHandel. "Introduction to the special issues on corpus methods." *Music Perception: An Interdisciplinary Journal* 31, no. 1 (2013): 1-3; Shanahan, Daniel, John Ashley Burgoyne and Ian Quinn, eds. *Oxford Handbook of Music and Corpus Studies*. Oxford: Oxford University Press, 2022.
[8] Casey, Michael A., Remco Veltkamp, Masataka Goto, Marc Leman, Christophe Rhodes, and Malcolm Slaney. "Content-based music information retrieval: Current directions and future challenges." *Proceedings of the IEEE* 96, no. 4 (2008): 668-696; Celma, Òscar, and Paul Lamere. "If you like Radiohead, you might like this article." *AI*

A particularly noteworthy application of MIR technology to the musicology of recorded music is to examine trends over chronological time through date-annotated musical audio files. The Million Song Dataset, a very large corpus of audio files pre-analyzed by the authors to avoid copyright issues in its distribution,[10] has been utilized to study chronological trends in Western popular music.[11] In popular music analysis, the hand crafted database of Billboard hits used by de Clercq and Temperley gave stronger evidence for the importance of the plagal cadence in popular music above the perfect.[12] Gauvin's 700 transcriptions of 1960s pop songs demonstrate an increase in 'flat-sided' harmony during the decade and thus the increasing disposition towards the Aeolian mode;[13] North and collaborators compare the commercial music markets in the UK and the US in terms of some proprietary algorithms for the detection of emotional state, working with 42714 pieces.[14]

We bring these threads together in an automatic computer analysis across 175 released Radiohead songs, covering all the major album releases, singles and b-sides up to the 20th anniversary *OK Computer* retrospective of 2017. In the majority of the text we exclude the 2017 remaster, and work with 163 songs including independent singles and albums from *Pablo Honey* (1993) up to *A Moon Shaped Pool* (2016).

The present study by design does not treat sociological aspects to address the more complete picture evoked by Walser,[15] but may still supply new information that can inform discussion of the affect, notability and value of Radiohead's work within culture. The advantage of the computer is to examine musical aspects across a large set of recordings, in relatively neutral fashion up to the programmers' transparency concerning their models.

---

*Magazine* 32, no. 3 (2011): 57-66; Lerch, Alexander. *An Introduction to Audio Content Analysis*. Chichester: Wiley, 2012; Müller, Meinard. *Fundamentals of Music Processing*. Springer, 2015.
[9] Cooper, Matthew, and Jonathan Foote. "Summarizing popular music via structural similarity analysis." In 2003 IEEE Workshop on Applications of Signal Processing to Audio and Acoustics, New Paltz. NY, 2003.
[10] Bertin-Mahieux, Thierry, Daniel P W Ellis, Brian Whitman, and Paul Lamere. "The Million Song Dataset." In *Proceedings of the 12th International Society for Music Information Retrieval Conference*, Miami, FL, 2011.
[11] Serrà, Joan, Álvaro Corral, Marián Boguñá, Martín Haro, and Josep Ll Arcos. "Measuring the evolution of contemporary western popular music." *Scientific reports* 2, no. 1 (2012): 1-6; Percino, Gamaliel, Peter Klimek, and Stefan Thurner. "Instrumentational complexity of music genres and why simplicity sells*." PloS one* 9, no. 12 (2014): e115255; Mauch, Matthias, Robert M. MacCallum, Mark Levy, and Armand M. Leroi. "The evolution of popular music: USA 1960–2010." *Royal Society open science* 2, no. 5 (2015): 150081.
[12] de Clercq, Trevor and David Temperley. "A corpus analysis of rock harmony." *Popular Music* 30, no. 1 (2011): 47-70.
[13] Gauvin, Hubert Léveillé. ""The Times They Were A-Changin": A database-driven approach to the evolution of musical syntax in popular music from the 1960s." *Empirical Musicology Review* 10, no. 3 (2015): 215-238.
[14] North, Adrian C., Amanda E. Krause, Lorraine P. Sheridan, and David Ritchie. "Comparison of popular music in the United States and the United Kingdom: Computerized analysis of 42,714 pieces." *Psychology of Music* 48, no. 6 (2020): 846-860.
[15] Walser, Robert. "Popular music analysis: ten apothegms and four instances." In *Analyzing Popular Music.* Edited by Alan Moore. Cambridge: Cambridge University Press, 2003, 16-38.

## 2 Technical details

The basis of an audio content based MIR system is some sort of signal processing engineering to derive musically pertinent audio features (also sometimes called descriptors). These features are typically of a reduced sampling rate than the original source audio, and correspond to low-level properties of sound such as the loudness, brightness, instantaneous predominant pitch, attack etc. in short windows of time, or higher level properties over longer time windows, such as a local chord, key, rhythmic or metrical structure. All such features can be subject to statistical analysis (such as a mean over a number of seconds or a whole piece), and sets (vectors) of such features at a given window in time compared to others (how do the different parts of a song, or different songs on an album, or different works across a whole corpus, relate?). The feature data is often input to machine learning algorithms such as classifiers (detect sound vs silence, detect speech vs music, detect one style vs one or more others, etc), or clustering (how do a set of examples group?).[16] In some cases, and especially as features become more high level, there may be what a human analyst would consider mistakes in the machine analysis when compared to a human transcriber, though MIR algorithms continue to improve in effectiveness (and this study could be re-run when feature extraction capabilities become available). The advantage of the machine is its untiring and neutral approach in tackling large amounts of audio data, and the open and precise specification of the features.

In the present work, 41 audio features were extracted; Table 1 provides a list of these and attempts to provide natural language description. Further details on the technical details of implementation using the open source software SCMIR are available (https://github.com/sicklincoln/SCMIR).[17] Each feature was sampled at a rate of just over 43 samples per second (corresponding to 44.1KHz audio and a hop size of 1024 samples); some features themselves depended on a longer working memory of audio, on the order of a few seconds (for instance, beat tracking decisions rely on examining some seconds worth of audio data and can't be calculated only on an instantaneous spectral frame). The 41 features are primarily timbral and rhythmic descriptors. Full pitch information, such as the accurate automatic extraction of all appearing melodic and bass lines, is beyond the state of the art in computer audio transcription.[18] However, automatic chord detection is relatively successful, and a specialist plugin[19] was also utilized to find chords within every piece in a separate analysis (see Section 4).

The selection of these features emphasized those that had a perceptual connection, such as loudness based on an auditory model, but had to take advantage of what was available already in software. There is a pragmatism to their selection; there are not equal numbers of timbral or rhythmic descriptors, which is a potential implicit issue when combining the features en masse, though the overall use of these features is consistent between pieces analyzed. It is readily acknowledged that perceptual studies have not been carried out to validate these, and only limited work in MIR has addressed such issues,[20] with no open source code implementations of fully perceptually validated features available (in contrast, all the features detailed here are available for researcher perusal and re-use in SCMIR). Nonetheless, it is valuable to see where current MIR methods can take us, understanding that at a later point we may revisit conclusions as more perceptually convincing feature detection algorithms arise.

---

[16] Casey et al. 2008; Lerch 2012; Müller 2015.
[17] Collins, Nick. "SCMIR: A SuperCollider Music Information Retrieval Library." In *Proceedings of the International Computer Music Conference*, Huddersfield, 2011.
[18] Benetos, Emmanouil, Simon Dixon, Dimitrios Giannoulis, Holger Kirchhoff, and Anssi Klapuri. "Automatic music transcription: challenges and future directions." *Journal of Intelligent Information Systems* 41, no. 3 (2013): 407-434.
[19] Mauch, Matthias, and Simon Dixon. "Approximate Note Transcription for the Improved Identification of Difficult Chords." In *Proceedings of the 11th International Society for Music Information Retrieval Conference*, Utrecht, The Netherlands, 2010.

**Table 1: 41 audio features extracted over the corpus**

| Feature number | Feature | Description |
| --- | --- | --- |
| 0 | Loudness | Psychoacoustic model of loudness |
| 1 | Sensory dissonance | Psychoacoustic sensory dissonance model after Sethares (2005) |
| 2-4 | 3 Energy bands | Energy for low (400Hz cutoff), mid (centred 3000Hz), and high frequency (cutoff 6000Hz) regions |
| 5-16 | Mel Frequency Cepstral Coefficients | Features much used in MIR systems, useful for timbral discrimination (they work particularly well at differentiating vowel sounds in speech recognition) |
| 17 | Spectral centroid | Measure of brightness of a sound |
| 18 | 50% Spectral percentile | Frequency below which 50% of the spectral energy falls |
| 19 | 90% Spectral percentile | Frequency below which 90% of the spectral energy falls |
| 20 | Spectral flatness | Measure of flatness of the spectral power distribution |
| 21 | Jensen-Shannon divergence | Compare all captured spectral distributions within the last two seconds; acts as a spectral change detector (so, similar spectral frames mean little divergence) |
| 22 | Spectral entropy | Entropy of the spectral power distribution |
| 23 | Transientness | Measure of transient energy in the signal, based on a wavelet transform |
| 24 | Harmonicity | Root mean square amplitude (over 1024 sample windows) of tonal (harmonic) component of signal after median source separation |
| 25 | Percussiveness | Root mean square amplitude (over 1024 sample windows) of percussive component of signal after median source separation |
| 26-28 | Onset statistics of percussive part of signal | In the last two seconds, the density (raw count) of attacks, and the mean and standard deviation of inter-onset intervals |
| 29-32 | Beat statistics of percussive part of signal | Beat histogram statistics; the entropy of the beat histogram, the ratio of the largest to the second largest entries in the beat histogram, the diversity (Simpson's D measure) of beat histogram, and metricity (consistency of high energy histogram entries to integer multiples or divisors of strongest entry) |
| 33-35 | Onset statistics of whole signal | As above, but calculated on the whole signal without any source separation step |
| 36-39 | Beat statistics of whole signal | As above, but calculated on the whole signal without any source separation step |
| 40 | Tempo | Estimated tempo of piece at a given moment in time |

Two figures are now presented to illustrate the audio feature analysis. Figure 1 plots three features in parallel for 'Creep' (1993), Radiohead's equivalent of Rachmaninov's 'C# minor prelude': an early work responsible for wider fame and often recalled by audiences despite the originator's unease. The three features are the perceptual loudness, the brightness, and the density of percussive onsets (0, 17 and 26 in Table 1), and the sampling takes an average over two second windows once per second. The two main choruses, with their heavy sustained distorted guitar chords, are obvious in the top trail. They are represented in the middle trail by relatively stable average brightness around 3500Hz and corresponding to the held distorted power chords. The peaks ahead of the chorus do not correspond to the notorious mid to low range 'chunky' lead guitar hits, but cymbal work by the drummer, which also has the effect of temporarily reducing the event rate detected. The influence of percussion is seen in greater detail if features are examined at a higher resolution, and the averaging effects necessarily to make global plots like this should be borne in mind (the top plot shows some bar by bar fluctuation but doesn't indicate individual instantaneous strikes).

---

[20] Friberg, Anders, Erwin Schoonderwaldt, Anton Hedblad, Marco Fabiani, and Anders Elowsson. "Using listener-based perceptual features as intermediate representations in music information retrieval." *The Journal of the Acoustical Society of America* 136, no. 4 (2014): 1951-1963.

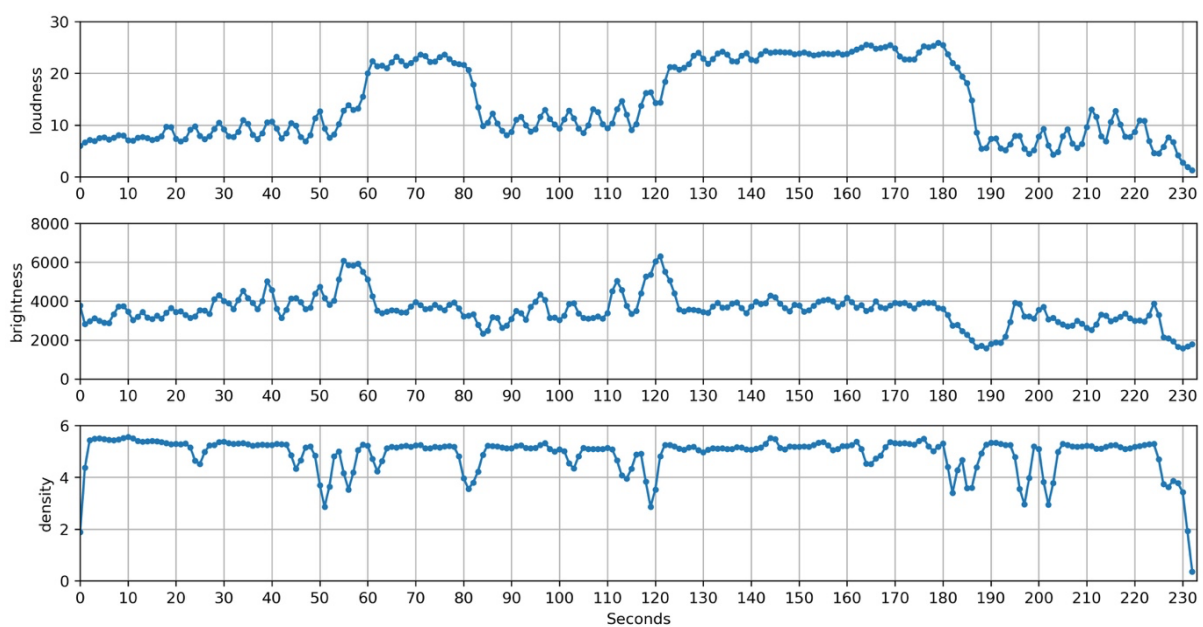

**Figure 1: Three feature tails over Radiohead's 'Creep' (1993). From top to bottom, the three features here are the perceptual loudness, the brightness, and the density of percussive onsets**

Figure 2 compares the three features of Figure 1 between the original CD of *OK Computer* (1997) and the remaster (2017), providing also a sanity check on the consistency of feature extraction. The instantaneous features are processed as means over each song on the album. The original mastering predates the peak of the 1990s to millennium 'loudness war' and the remaster was mastered by Bob Ludwig, a vocal critic of excessive mastering volume.[21] There is a very close relationship across the two versions of the album showing high consistency; the Spearman's correlations (which do not assume any Gaussian distribution of values) for each subplot from the top to the bottom are 0.972, 0.993 and 0.944. However, it is apparent that Ludwig's 2017 remaster (dashed line) is a little louder (notwithstanding Ludwig's loudness criticism), and a touch less bright. These slight differences point to the variability that may be found between different releases of the same albums, with the difficulty of differentiating in audio files artistic decisions by the band from production line decisions such as mixing or mastering. A larger scale study of remastered versions of albums across many artists may illuminate this more.

---

[21] Owsinski, Bobby. *The Mastering Engineer's Handbook*. Burbank, CA: Bobby Owsinski Media Group, 2017.

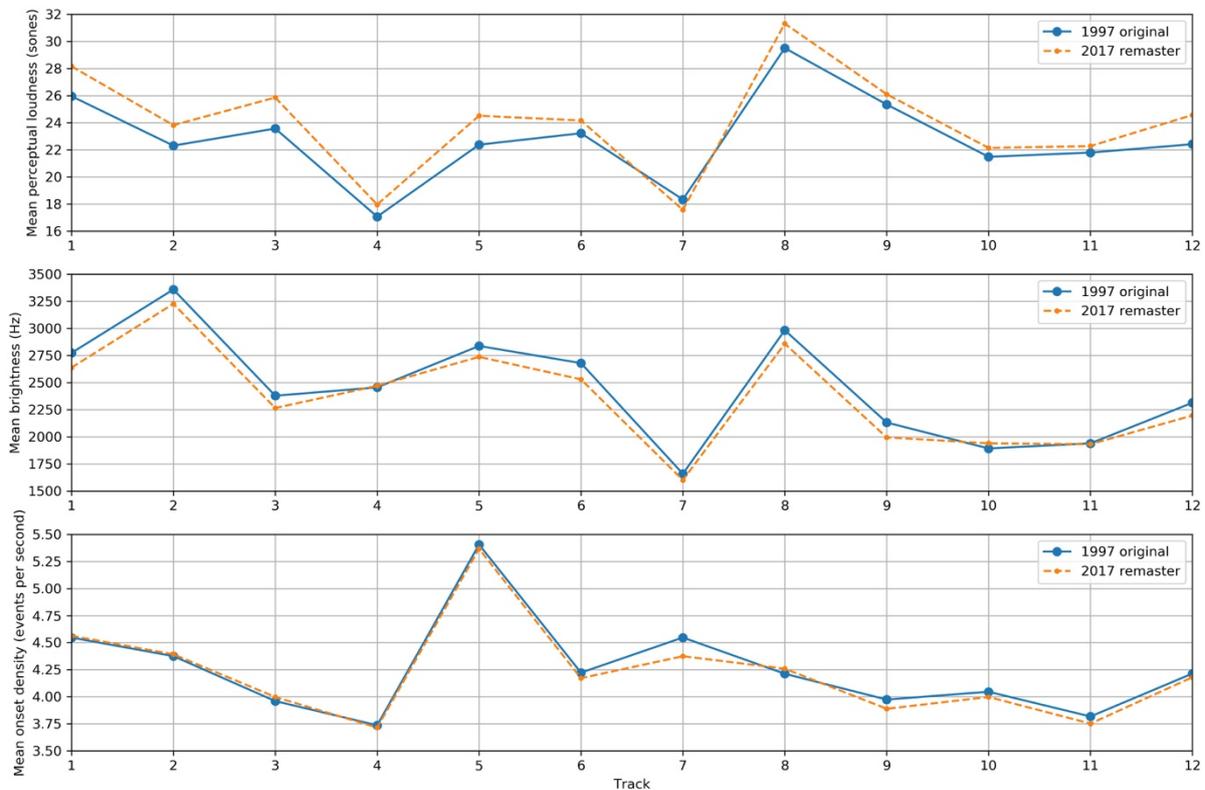

**Figure 2: Comparison of the original CD of *OK Computer* (1997) and the 2017 remaster, with respect to three features over the 12 album tracks (in album order from 1. 'Airbag' to 12. 'The Tourist')**

**3 Chronological analysis from timbral-rhythmic features**

If a set of audio files have associated date information, such as a commercial release date or studio recording session completion date, feature data from analysis of those files can be examined in chronological order. The year of release is the most straight forward to apply and used here. Release years for Radiohead were obtained from the physical media of the releases whenever possible, and otherwise double checked or obtained for album independent download releases through online sources (discogs.com and https://en.wikipedia.org/wiki/Radiohead_discography). As Figure 3 shows, there are years without any associated release, and years corresponding to album release or singles.

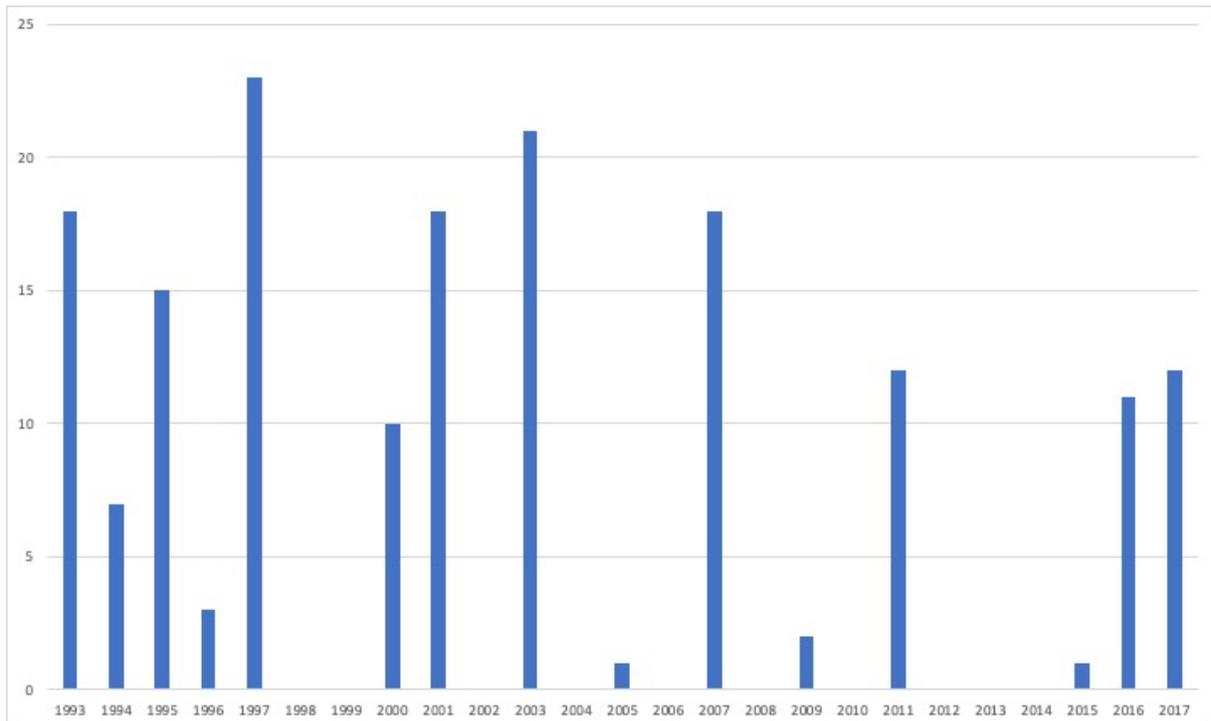

**Figure 3: Radiohead releases by year in the corpus**

### 3.1 Time variation of features

Songs can be plotted against time via summary features, such as an average over the whole track. A scatter plot can show the values for all songs for a given feature chronologically; we can then fit via linear regression a trendline to examine if there is any statistically significant trend over time, though there is a proviso that explained variance isn't necessarily high. Figure 4 plots four features averaged by song, the four which had the highest significance scores within the corpus; the fitting and significance assessment was conducted with Python's scipy library, namely, the stats.linregress function. All four are actually related, in that they show an increase in low frequency energy, and a drop in brightness, with associated reduction in spectral entropy ('spectral envelope noisiness'), over Radiohead's releases. An immediate explanation is the drop in the use of overtone heavy guitar distortion from their early career, and an increase in synthesized bass lines, or bass pushed up in mixes. None of the four have coefficient of determination ($r^2$) above 30%, showing the weakness of a linear model in explaining the year by year variation (to 3 d.p., for feature = 2 r = 0.542 p = 8.028e-14 $r^2$ = 0.294, feature = 18 r= -0.504 p = 6.83e-12 $r^2$ = 0.254, feature = 22 r = -0.475 p = 1.558e-10 $r^2$ = 0.225, feature = 17 r = -0.458 p = 7.939e-10 $r^2$ = 0.2097).

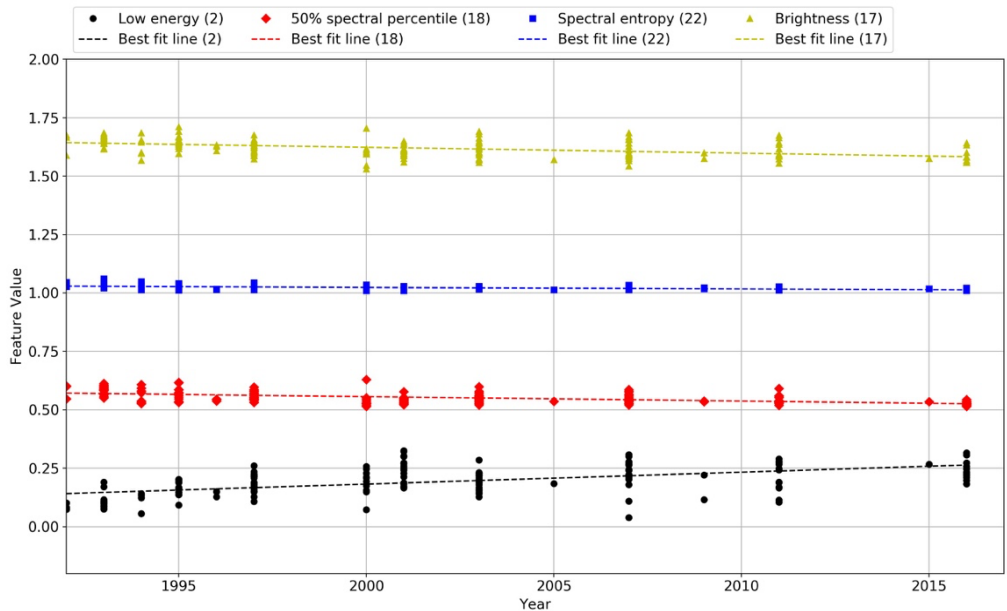

**Figure 4: Feature averages per song for four features; the legend gives each feature name and the associated number in table 1 above. Note that the original feature values between 0 and 1 have been offset by 0.5 between each trail, to facilitate a shared plot without overlap**

Two further feature trails are worth explicitly noting. The loudness and the tempo over time are plotted in Figure 5 to investigate a potential 'ageing' effect for a rock band,[22] who might mellow out over many albums (feature 0 (loudness): r = 0.262 p = 0.00074175 $r^2$ = 0.0685 feature 40 (tempo): r = -0.224 p = 0.0041 $r^2$ = 0.05). We see a fall off in tempo over time, though a jump back up for *A Moon Shaped Pool* (2016). The tempo information here derives from a computational beat tracker biased to select a metrical level closest to 100-120 bpm; if a track is particularly slow, leading to the beat tracker counting at double the rate, or no consistent tempo is found, the average tempo may jump back up again. The loudness of the album mix increases over time however, in line with the general trend for mastering in popular music. The explained variance is low, however, so such results should not be taken as particularly robust.

---

[22] Puri, Samir. "Cycles of metal and cycles of male aggression: Ageing and the changing aggressive impulse." In *Can I Play with Madness? Metal, Dissonance, Madness and Alienation.* Edited by Colin McKinnon, Niall Scott, and Kristen Sollee. Leiden: Brill, 2011, 101-109

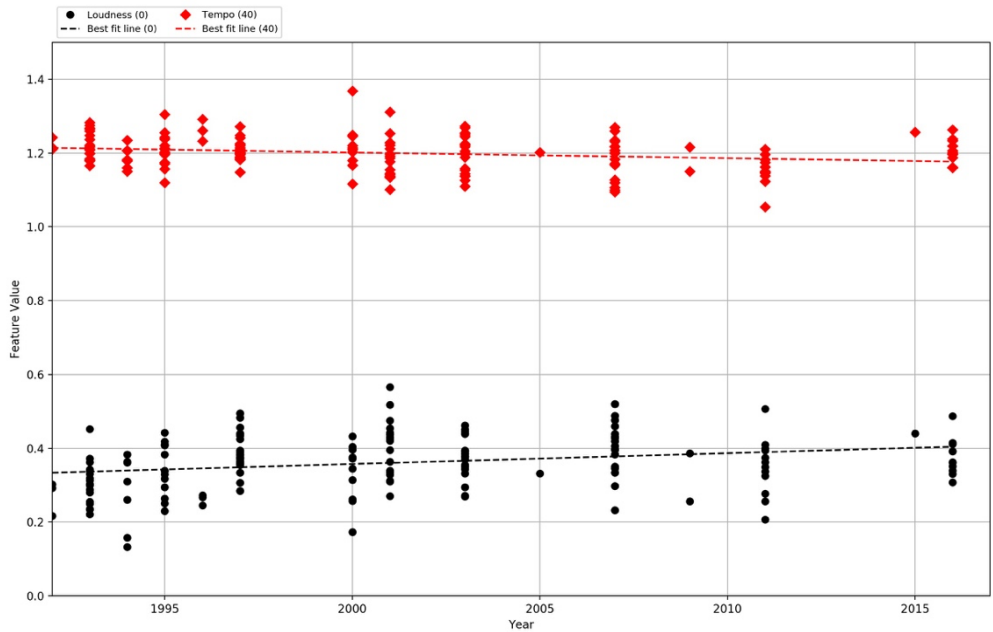

**Figure 5: Feature averages per song for loudness and for tempo; the legend gives each feature name and the associated number in table 1 above. Note that the original feature values between 0 and 1 have been offset by 1 between each trail, to facilitate a shared plot without overlap**

### 3.2 Chronological similarity matrices

In Music Information Retrieval the similarity matrix[23] is often used to examine structure within an individual song, but applicable also to assess similarity between songs, albums, genres, years, and more. The self-similarity matrix gathers all pairwise comparisons between different segments of a song with itself, or the similarity matrix more generally gathers all comparisons between different entities in a set. The matrix is necessarily symmetric when the similarity measure *d* does not depend on the order of consideration of two entities x and y (*d*(x,y) = *d*(y,x)). A similarity matrix is critically dependent on the choice of features to represent each entity to be compared, and the metric used for comparison of the feature vectors. Whether to use all features, or a particular subset, may have a strong effect on the results. To acknowledge this, we compare six different feature subsets to form six similarity matrices over the nine main album releases, as plotted in figure 6. Each row and column is labelled with a code indicating one of the 9 albums under discussion (see Figure caption). The matrix of comparisons is symmetric since the distance of album a to b is the same as that from b to a. The whiter the monotone, the closer the similarity; the diagonal, of the albums compared to themselves, is maximally bright for being exactly the same.

      The distances used for calculation here are based on the standard Euclidean distance between the average feature vectors per song, or per album (the city block and cosine distances were also tried informally but didn't lead to any immediate advantage). So if using all features, similarity is the square root of the summed squared differences over 41

---

[23] Cooper and Foote 2003; Casey et al. 2008.

dimensions, e.g. over the 41 features in Table 1, with all values max-min normalized to a common 0 to 1 range. The max-min norm may be prone to outliers, though does not make any assumptions about Gaussian distribution of data. Because the 41 features are predominantly short-term timbral and rhythmic descriptors, notions of similarity here prioritize average timbre and groove over higher level facets of song writing and vocal and instrumental expression. Other measures might be substituted, such as models trained on a particular song or album and used to predict other songs or albums, with the ease of prediction a measure of similarity (how well does a given song 'explain' another one?).

A research question is whether the similarity matrix data supports the narrative that Radiohead made large changes for *Kid A* (2000), and after its companion album *Amnesiac* (2001) from related sessions, returned to their core guitar-led composition for subsequent albums, even if continuing to utilize an *OK Computer* (1997) style extended timbral palette. The matrix for all 41 features (top left) demonstrates a link between *Hail to the Thief* (2003) and *OK Computer*, a big change from *OK Computer* to *Kid A* and the closest move between albums being that from *The Bends* (1995) to *OK Computer*. All of the matrices apart from the bottom right have proximity of *Kid A* and its sister album *Amnesiac*, though the harmonicity feature on its own (bottom right) shows a different association, with a grouping of *The Bends*, *OK Computer* and *Kid A*. The lower left matrix is from the most discriminative single feature, the 90% spectral percentile, the feature that led to the greatest mean differences between albums of any of the 41. This single feature supports the narrative of the *OK Computer* to *Kid A* jump, the *Kid A* and *Amnesiac* proximity, and the links between *OK Computer* and *Hail To The Thief*. It also posits a closeness of *In Rainbows* (2007) and *The King of Limbs* (2011), and the difference of *A Moon Shaped Pool* (2016) to the other 8 albums. Nonetheless, in musical terms the percentile is a proxy of brightness of mix, which may be correlated with levels of guitar distortion, percussion and the like; it makes sense that *A Moon Shaped Pool*, as perhaps the most mellow of the Radiohead albums, stands out by this measure.

**Figure 6: Similarity matrices over nine Radiohead albums: PH *Pablo Honey* (1993) TB *The Bends* (1995) OKC *OK Computer* (1997) KIDA *Kid A* (2000) AMN *Amnesiac* (2001) HTTT *Hail to The Thief* (2003) IR *In Rainbows* (2007) TKOL *The King of Limbs* (2011) AMSP *A Moon Shaped Pool* (2016). The six feature sets used with indices referring to Table 1 are: all features (top left), rhythm features 26-39 (top right), spectral timbral features 17-22 (middle left) , MFCCs 5-16 (middle right), 90% spectral percentile 19 (bottom left), harmonicity 24 (bottom right). The actual distances are plotted above each shaded square; the lightness of shading represents the similarity (smaller distances are brighter, larger distances are darker).**

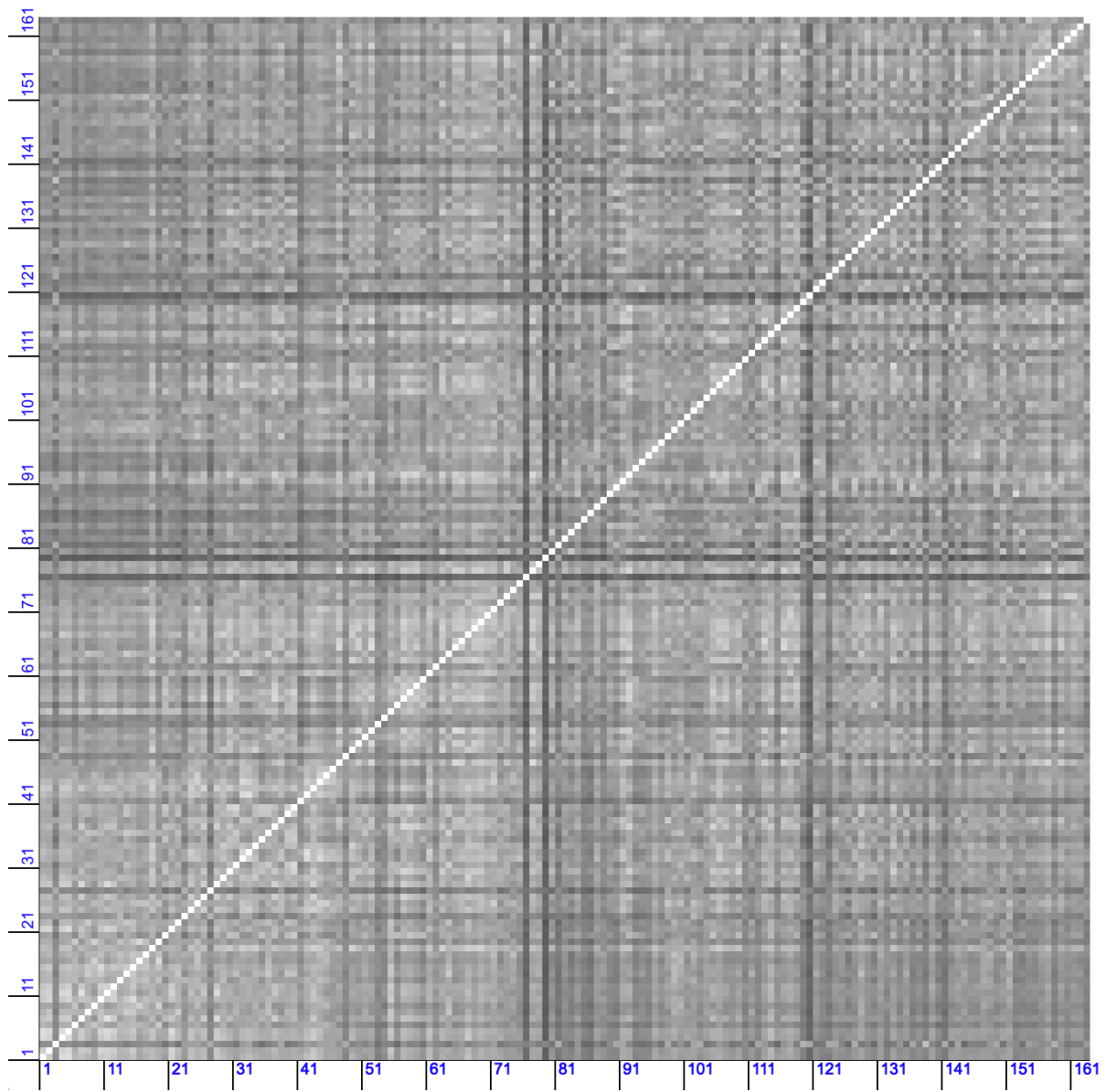

**Figure 7: Similarity matrix over 163 Radiohead songs from 1992-2015. The maximal self-similarity diagonal is clear, but the structure is otherwise complicated, and not obviously self-contained by album groupings except for a slightly lighter more similar area at the bottom left. Maximally dissimilar songs appear as horizontal and vertical (symmetric) darker strips.**

Figure 7 plots the matrix per song, with songs ordered chronologically, over the 163 songs not including the remastered 2017 *OK Computer* main disc, with respect to all 41 features. This matrix is much harder to interpret, since there is substantial variation within the songs of an album, let alone in comparison to all other albums and album-independent releases. It does demonstrate many potential heterarchical links between time periods. There is a slightly brighter bottom left square corresponding to the early Radiohead material up to and including *The Bends* (1995).

We also note the opportunity to find the ten most representative Radiohead songs, and the ten strangest outliers. Table 2 was constructed by summing the columns of the song similarity matrix, where those songs maximally dissimilar to all others will score as most dissimilar overall, or vice versa for similarity.

**Table 2 Similarity scores for most representative (most typical) to least representative (least typical) Radiohead songs.**

| Similarity rank | Dissimilarity score (lower is more similar to all other songs in the corpus) | Song |
| --- | --- | --- |
| 1 | 7.869951 | 'The Tourist' |
| 2 | 8.130662 | 'Banana Co' |
| 3 | 8.145942 | 'Optimistic' |
| 4 | 8.219033 | 'Polyethylene, Pt. 1 & 2' |
| 5 | 8.242041 | 'Subterranean Homesick Alien' |
| 6 | 8.315117 | 'Ful Stop' |
| 7 | 8.317736 | 'Cuttooth' |
| 8 | 8.356912 | 'A Reminder' |
| 9 | 8.483929 | 'No Surprises' |
| 10 | 8.5043 | 'Karma Police' |
| … | … | … |
| 154 | 13.35455 | 'Like Spinning Plates' |
| 155 | 13.417979 | 'India Rubber' |
| 156 | 14.2353 | 'Harry Patch (In Memory Of)' |
| 157 | 14.391337 | 'You Never Wash Up After Yourself' |
| 158 | 14.473617 | 'Codex' |
| 159 | 14.595802 | 'Hunting Bears' |
| 160 | 15.001918 | 'MK 2' |
| 161 | 18.775993 | 'MK 1' |
| 162 | 20.109289 | 'Motion Picture Soundtrack' |
| 163 | 20.674503 | 'Treefingers' |

Listening to this, and the wider Radiohead corpus, does demonstrate the efficacy of the technique, at least as far as general song timbre goes. The outliers are stranger choices of sound world in the Radiohead corpus; the most average songs are those utilising Radiohead's core instrumentation of guitars, drums and Yorke's voice. Their electronica experiments do not make enough of a dent to be deemed representative, on average, and they remain at the core a guitar band, albeit one quite willing to explore alternative instrumentations during their career.

**3.3 Prediction of album from a song, and of the year of release of any Radiohead song**

Supervized machine learning algorithms can be applied to the task of predicting the year of release of a given song in the corpus, or which album contains a given song. We associate each song with its local average feature vectors (more complicated models are certainly possible) to give a baseline for this sort of prediction. The success of prediction gives clues as to how easily the songs at given times or for given albums are distinguished by their timbral-rhythmic signatures. There may be a potential confound with the cohesion brought about by mixing and mastering a given release, but human listening can certainly distinguish larger compositional shifts over the band's career, and it is interesting to see how well a machine can do at distinguishing different eras and releases.

To discriminate the nine main Radiohead albums (see Figure 6), a single layer neural net with 41 inputs, 41 hidden units, and 9 outputs (one for each album) was trained over 1000 epochs. A 50/50 split of randomly allocated training and test data was used; the data originated from two second averages of the 41 features used in this article, taken at one second hops throughout songs. Segments from the same song were only allowed into either the training or the test set (if the split did not differentiate songs, results were much improved but less representative of the actual level of machine performance). There were 12750 training instances, for which the trained neural net classified 3979 correctly, giving a success percentage of 31.208% (note that picking randomly has a 1/9 = 11.111% chance of success). Over the unseen test set, the trained net performed correctly 3214 out of 12850 times, 25.012%. This is a weakly generalising performance for distinguishing 9 classes, though not quite as poor as chance. It was possible to improve with a NaiveBayes model with 9 outputs and 41 inputs, obtaining for a random (song preserving) training set 5467 out of 12750 instances correct (42.878% success rate) and for the test set 3871 out of 12567 correct (30.803%).

To discriminate the fifteen years in which Radiohead actually released songs from 1992 to 2016 (see Figure 3) , a single layer neural net with 41 inputs, 41 hidden units, and 15 outputs (one for each possible year) was trained over 1000 epochs. A new randomly allocated 50/50 split of training and test data was used (again avoiding allocating any segments from the same song to both training and test sets, that is, a given song only appears in one of the training or the test sets). There were 18536 training instances, for which the trained neural net classified 3745 correctly, giving a success percentage of 20.204% (note that picking randomly has a 1/15 = 6.667% chance of success). Over the unseen test set, the trained net performed correctly 2084 out of 20287 times, 10.273%. This is a weak generalising performance for distinguishing 15 classes, showing that the computer finds this a hard task, and that the musical sonic features that distinguish different parts of the band's career are not well captured by the data, if we assume that there is sufficient variation in their career to distinguish their work over time.

A Naïve Bayes model was again used to try to improve matters, this time in combination with a greedy feature selection algorithm (at each step, add the best performing feature from those not yet used, keeping track of the best overall results across all stages). Because the Naïve Bayes algorithm couldn't cope with single examples for a year, three classes were created splitting the years of release into three groups ( first 1992, 1993, 1994, 1995, 1996, 1997, that is, up to and including OK Computer, second group the Kid A/Amnesiac years 2000 and 2001, and third group 2003, 2005, 2007, 2009, 2011, 2015, 2016). The best scoring model used fourteen features (numbers 2, 3, 5, 6, 8, 10, 12, 15, 21, 22, 25, 29, 32 and 37), obtaining over 18858 training examples 59.200% correct and over 20128 test instances 67.891% accuracy. With chance at 33.3r% this is reasonable performance, though contrived to three eras of songs rather than 15 distinct years of releases.

A human expert Radiohead fan would score very highly at recognising portions of songs and assigning them to albums and years of release. The computer with the machine learning algorithms and feature data here is by no means successful, though does perform better than chance. This suggests at the very least an objective baseline as to how successfully Radiohead's oeuvre is distinguishable. The more machine results can be improved, the greater the evidence that the band have not rested on their laurels and have successfully varied the timbral-rhythmic features of their music over their career.

## 4 Harmonic analysis

Another perspective on the Radiohead corpus is provided via harmonic analysis. Although Brad Osborn identifies many interesting pitch structures and voice leading techniques within Radiohead's music,[24] he works by selected examples, and not across all available pieces where the computer can give us a different angle of analysis. A mid level musical understanding task which computers can perform relatively well at is chord transcription. Reliable and consistently formatted MIDI files or other lead sheet data was not available for the song corpus, and was beyond the scope of this study to establish; it was convenient to operate on the same set of audio data as previous sections directly.

The 163 Radiohead songs from 1992-2016 were analyzed via the vamp plugin Chordino (http://www.isophonics.net/nnls-chroma) as developed by Matthias Mauch,[25] which extracts time-stamped chord labels from audio files. Custom post processing was used to extract and analyze the chord sequences. Chords were classified according to ten chord types: major, minor, dominant $7^{th}$, major $7^{th}$, minor $7^{th}$, diminished, major $6^{th}$, minor $6^{th}$, minor $7^{th}$ b5, augmented. The minor $6^{th}$ and the minor $7^{th}$ flattened $5^{th}$ are related by inversion; this set constitutes the baseline for the types extracted by Chordino. We did not, however, take account of some rare cases of chords over alternative roots, and Chordino occasionally outputs an 'N' if it cannot identify a chord; such results were excluded from subsequent analysis rather than form an additional category. The analysis here is at the mercy of the effectiveness of the Chordino plugin, which performs competently within the space of chord detection by machine, though not at the level of a human analyst, and with the assumption that the chord types detectable exhaust those appearing; we openly admit Osborn 2017 presents a richer field of harmonic models. A further analysis of assignment errors by Chordino is also beyond the present scope, and we proceed understanding that the analysis may be reworked if a yet more competent algorithm arrives in the future, or algorithms arise for alternative harmonic models than common practice rock harmony and major-minor key systems.

Aggregate data for chord roots, and chord types, is presented in Figures 8 and 9 respectively. The most common roots correspond with the simpler guitar and piano keys, and flat keys appear less often. The chord types favour plain major and minor rather than more elaborate 7ths, and major 7ths appear more often than dominant 7ths.

---

[24] Osborn 2017.
[25] Mauch and Dixon 2010.

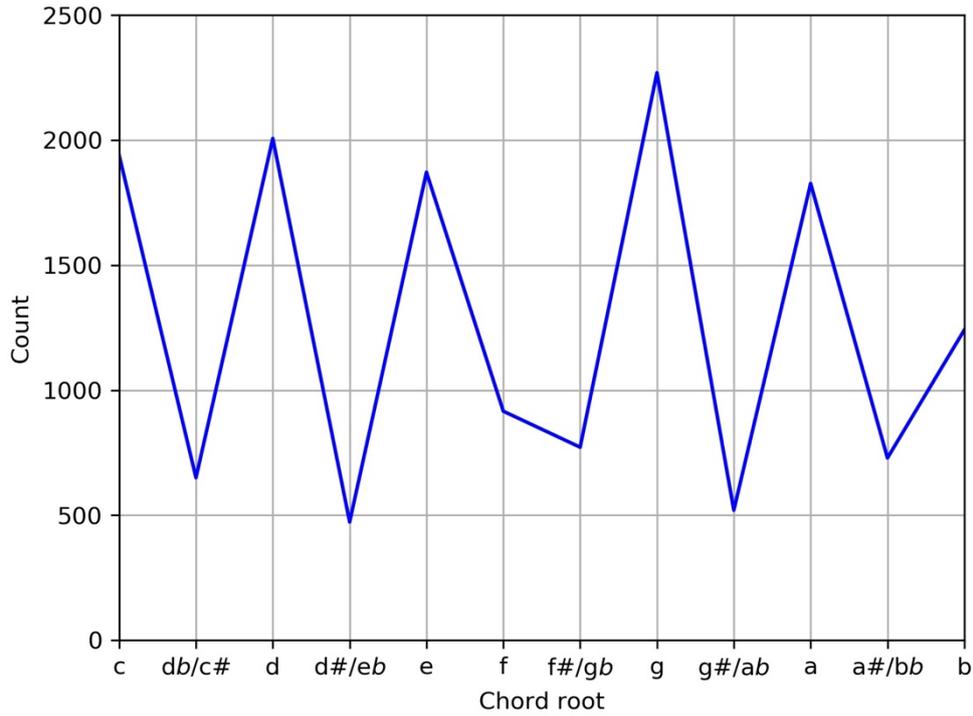

**Figure 8 Histogram over roots for all extracted chords in the 163 Radiohead songs 1992-2016**

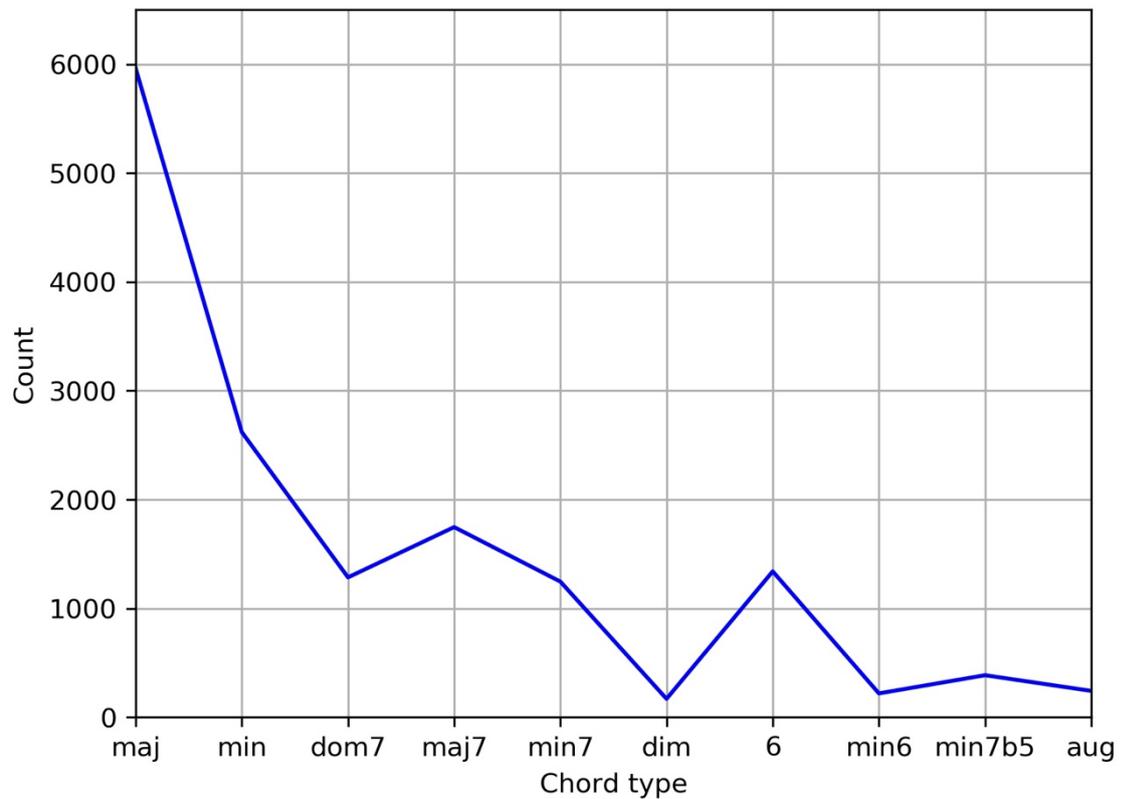

**Figure 9 Histogram over chord types for all extracted chords in the 163 Radiohead songs 1992-2016**

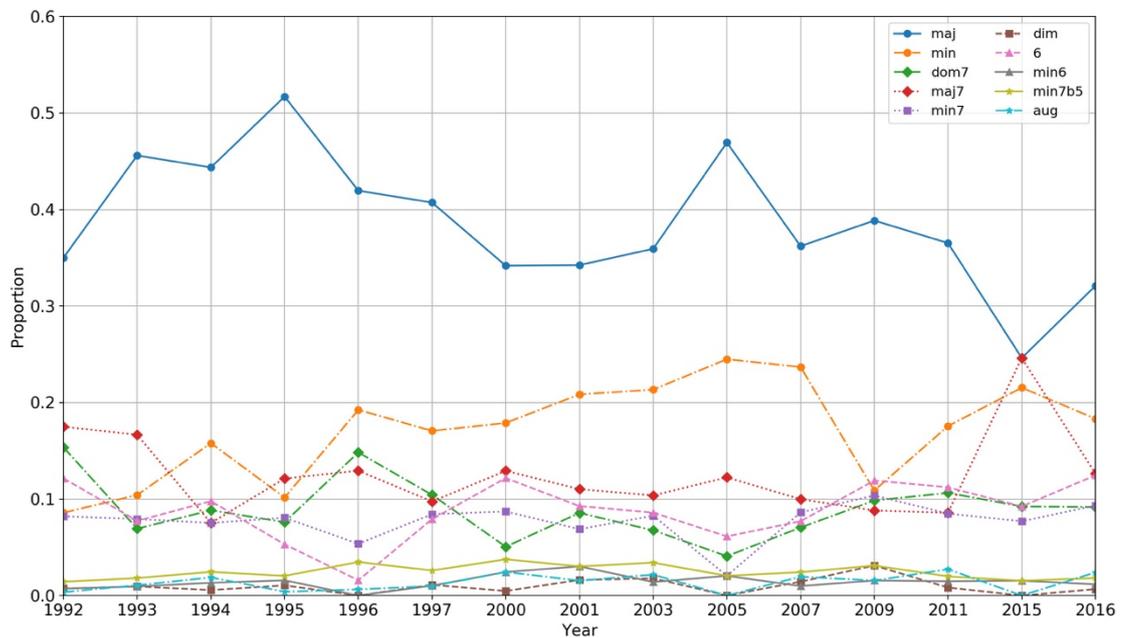

**Figure 10 Change in chord type usage by year**

Figure 10 supplies a breakdown by year of use of chord types, with one line in the diagram for each chord type. There is some fluctuation, such that the use of more complex chords is generally rarer, but minor chord use has a trend to increase over time (we resist the temptation to claim that Radiohead have become increasingly miserable over time).

In order to evaluate chord transitions, an additional automatic key extraction was run over the corpus to try to disambiguate functional harmony, with the results of the most common twenty transitions detailed in Table 3 below. Local key was determined by looking at five chords either side of a given chord (seven or even fifteen either side gave similar key estimations) , and finding the best matching major or minor key according to correlation with Krumhansl-Kessler profiles.[26] The most common transitions were identified by a histogram over 14400 possibilities using the mapping between two sets of 10 chord types * 12 chromatic roots (120 options at each stage of the bigram from chord to chord, so 12-*120 = 14400 transition types). Only transitions within the same key (both start and end chord were assigned the same major or minor key) were counted.

---

[26] Rowe, Robert. *Machine Musicianship.* Cambridge, MA: MIT Press, 2001

**Table 3 Most frequently occurring functional harmonic transitions within 163 Radiohead songs**

| Rank | Incidence Count (out of 10440) | Proportion | Start root | End root | Start type | End Type | Example Roman notation | Example (transposed to C) |
|---|---|---|---|---|---|---|---|---|
| 1 | 470 | 0.045 | 5 | 0 | major | major | IV->I | F->C |
| 2 | 401 | 0.0384 | 0 | 5 | major | major | I->IV | C->F |
| 3 | 149 | 0.0143 | 0 | 5 | major | major 7th | I->Imaj7 | C->Cmaj7 |
| 4 | 145 | 0.0139 | 5 | 0 | major 7th | major | IVmaj7->I | Fmaj7->C |
| 5 | 139 | 0.0133 | 0 | 0 | major | minor | I->i | C->Cm |
| 6 | 135 | 0.0129 | 5 | 0 | minor | major | iv->I | Fm->C |
| 7 | 134 | 0.0128 | 7 | 0 | major | major | V->I | G->C |
| 8 | 121 | 0.0116 | 0 | 5 | major | minor | I->iv | C->Fm |
| 9 | 120 | 0.0115 | 0 | 0 | minor | major | i->I | Cm->C |
| 10 | 113 | 0.0108 | 0 | 7 | major | major | I->V | C->G |
| 11 | 106 | 0.0102 | 9 | 0 | minor | major | vi->I | Am->C |
| 12 | 95 | 0.0091 | 0 | 0 | major | dominant 7th | I->I7 | C->C7 |
| 13 | 92 | 0.0088 | 8 | 0 | major | minor | VIb->i | Ab->Cm |
| 14 | 90 | 0.0086 | 5 | 0 | major 6th | major | IV6->I | F6->C |
| 15 | 87 | 0.0083 | 0 | 5 | minor | major | i->IV | Cm->F |
| 16 | 80 | 0.0077 | 0 | 3 | major | major 6th | I->iii6 | C->Eb6 |
| 17 | 75 | 0.0072 | 0 | 5 | major | major 6th | I->IV6 | C->F6 |
| 18 | 74 | 0.0071 | 5 | 0 | major | minor | IV->i | F->Cm |
| 19 | 74 | 0.0071 | 0 | 0 | major | major 7th | I->Imaj7 | C->Cmaj7 |
| 20 | 69 | 0.0066 | 0 | 8 | minor | major | i->VIb | Cm->Ab |

A prediction following De Clercq and Temperley's study[27] is that plagal (IV->I) chord movements should be much more frequent in recent popular music than perfect (V->I). Table 3 supports this for Radiohead, with plagal movement around 3.5 times more frequent than perfect. Also frequently appearing are major-minor alternations, and the move from minor to major chord such as iv->I, a frequent fixture in 1990s indie music in particular.

The chord data can supply additional information: inter-chord timings provide a measure of harmonic rhythm. The average gap between chord changes in seconds is plotted against year in Figure 11. A slowing over the career as the band age is readily apparent, and corroborates the data in Figure 5, which also showed an overall trend of slowing, albeit with a similar slight return to faster paced events in more recent releases. The peak for slow harmonic rhythm (maximum average chord gaps) appears earlier than the lowest average tempo of Figure 5, indicating some remaining independence of these aspects.

---

[27] De Clercq and Temperley 2011.

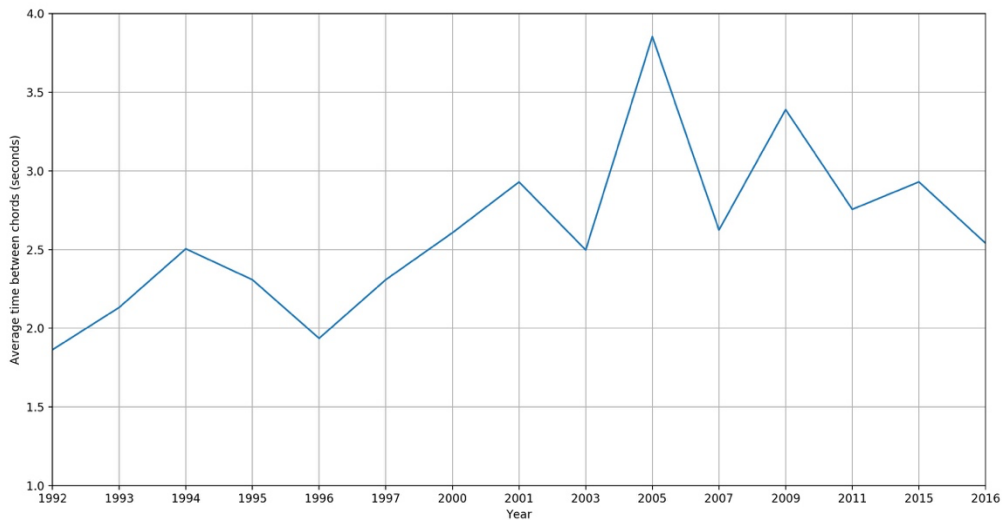

**Figure 11 Harmonic rhythm plotted as average time between chords over songs released in a given year**

## 5 Discussion

The analysis above is restricted to the work of Radiohead alone, and does not place the band in a wider context of other artists. The exploration of musical connections to other bands' recordings, and larger scale groupings of musical style, is a fertile area for further development. Indeed, the computational treatment of musical influence has already seen studies, for instance into the relationship amongst early synth pop albums,[28] sampled and sampling artists identified through whosampled.com[29] and within the aforementioned Million Song Dataset.[30] In the context of Radiohead, influencing bands and singers might include R.E.M., Joy Division/New Order, The Smiths, Siouxsie and the Banshees, The Pixies, Nick Drake, Neil Young, Scott Walker, Serge Gainsbourg and more, as revealed by the band themselves in interviews, through their choice of cover versions, and in critical and fan discussion around the band. Comparative analysis of recorded music careers amongst bands may be productive, for instance, in terms of spotting any more general mellowing of musical statements (e.g., dulling of timbre, slower tempi, less events per second), or exploration of alternative timbres and techniques over time. For instance, New Order's greater adoption of synthesizers and sequencers in the early 1980s might be set against Radiohead's exploration of expanded music technology peaking with *Kid A/Amnesiac*.

      The originality of any artist is contingent upon their upbringing in and reaction to wider culture. It is difficult for any analyst, whether human or proxy computer, to recover all

---

[28] Collins, Nick. "Computational Analysis of Musical Influence: A Musicological Case Study Using MIR Tools." In *Proceedings of the 11th International Society for Music Information Retrieval Conference*, Utrecht, The Netherlands, 2010.

[29] Bryan, Nicholas J. and Ge Wang. "Musical Influence Network Analysis and Rank of Sample-Based Music." In *Proceedings of the 12th International Society for Music Information Retrieval Conference*, Miami, FL, 2011.

[30] Shalit, Uri, Daphna Weinshall, and Gal Chechik. "Modeling musical influence with topic models." In *International Conference on Machine Learning*, Atlanta, USA, 2013.

the cultural influences upon a subject of their study at any given moment of time. A data set containing only Radiohead's music must necessarily miss outside influences; a larger data set enclosing the band corpus may still not be sufficiently broad to capture all precedents. Figure 12 provides an example. The top line in the figure is the guitar riff underlying Radiohead's 'Lucky' (1995), which outlines the underlying A major to E minor harmonic movement; the song is in E minor but with the major IV chord in the chorus (the motif loops longer with variants of these two bars, but the basic shape is illustrated). However, a similar harmonic movement and theme outlining IV->i appears in a contemporary piece, Björk's 'Isobel' (1995), in the main string riff (middle line). Both in turn share a precedent in The Eagles' 'Journey of the Sorcerer' (1975) guitar/string riff. Radiohead's prior contact with The Eagles example may have come more via *The Hitchhiker's Guide to the Galaxy* radio/TV shows (Thom Yorke has admitted the influence of the books, explicit also through the song title 'Paranoid Android'). These connections show that a more wide-ranging database of popular culture , even beyond a corpus of contemporary songs or lineage of 'indie/alternative' music, may be necessary to build up a more complete picture.

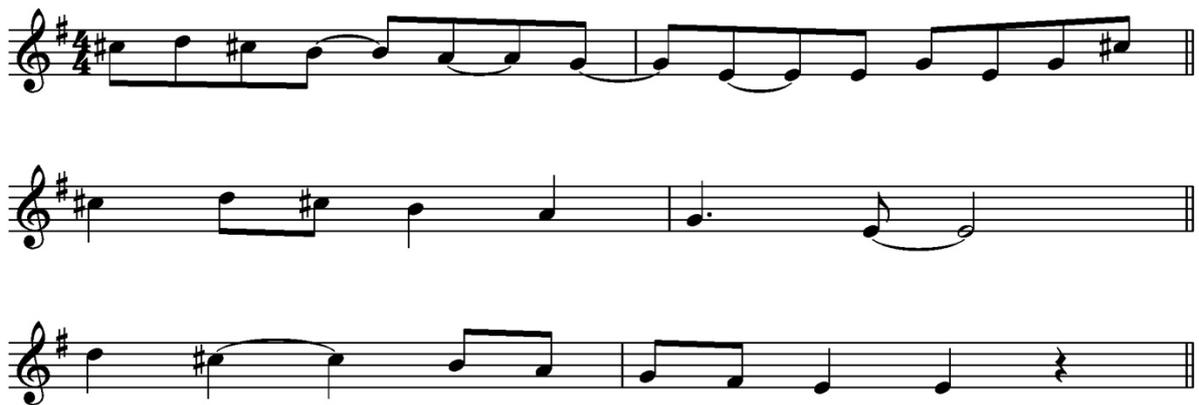

**Figure 12 Comparison of the chorus guitar riff for Radiohead's 'Lucky' (recorded 4 September 1995, top line) with Björk's 'Isobel' string riff (1995, middle line) and guitar line from The Eagles' 'Journey of the Sorcerer' (1975, bottom line). The two latter examples were transposed to E minor for ease of comparison.**

In this context, the putative results of Celma and Lamere on the most similar bands suggested through various forms of computational music recommendation[31] does not stand up completely to musicological scrutiny, since it may reveal proximity in a social media tag or audio feature space, but does not correspond to richer human experience, for instance, the listening and training histories of the band members themselves. There are many additional connections to find, separability not necessarily straight forward to access through audio. For example, The Farm's 'Groovy Train' (1990) includes the lyric 'you're so special' in direct prelude to the 'Creep' lyric 'so fucking special'; audio analysis of the music is not at a reliable level for automatic voice extraction and lyric transcription, even assuming both these songs were in a joint corpus, though a corpus of meta-data including lyrics would make this more accessible.  It would also be of interest to compare the band member's solo project music to

---

[31] Celma and Lamere 2011.

the main corpus, or to explore *MiniDiscs [Hacked]* (2019) as extensive demo material for *OK Computer.*

Nonetheless, the present study has implications for new directions in popular music studies founded in audio file signal analysis. MIR audio content analysis allows the examination of popular music recordings at a number of granularities:

1. The variation within a single piece
2. The diversity of pieces within an album
3. The variation of works by an artist over a timescale of years, including their complete career
4. The comparison of work between an artist and their peers, predecessors, and musical descendents

This form of analysis has contributions to make then from the analysis of individual works, to larger scale narratives over a band's output, the activity within a genre and even between genres for very large corpora. Future studies of the cultural evolution of popular music will inevitably draw further on such methods. There will remain qualifications on the level of machine listening obtained from computer analysis. Programming decisions can lead to different answers to research questions based on the choice of feature subsets and algorithms (compare Figure 6). As improved musical feature extraction and analysis techniques, more fully perceptually validated, become available for this research, the audio file corpus can be analyzed afresh to make further and increasingly robust discoveries. Deeper analysis can be applied to enrichen the results here, if bearing in mind the potential limitations of MIR algorithms and controversies of musical similarity, for example in a more systematic inter-album comparison based on multiple parallel models.